
%
%
%
%
%

\documentstyle{amsppt}
\nopagenumbers
\hsize 32pc
\vsize 50pc
\emergencystretch=100pt

\magnification =1200
\def\ms{{\medskip}}

\def\k{{\kappa}}
\def\t{{\tau}}

\def\slr{{\Cal R}}

\def\sls{{\Cal S}}

\def\slp{{\Cal P}}

\def\smalltype{\let\rm=\eightrm \let\bf=\eightbf
\let\it=\eightit \let\sl=\eightsl \let\mus=\eightmus
\baselineskip=9.5pt minus .75pt \rm}
\parindent=30pt

\def\myequal{\, = \,}
\def\onepsi{\hbox{\hbox{$^1$}\kern-.15em $\Psi$}}
\def\onejt{\hbox{\hbox{$^1$}\kern-.25em $\tilde J$}}
\def\onext{\hbox{\hbox{$^1$}\kern-.25em $\tilde X$}}
\def\oneit{\hbox{\hbox{$^1$}\kern-.25em $\tilde I$}}
\def\onert{\hbox{\hbox{$^1$}\kern-.1em $\tilde \slr$}}
\def\oneslr{\hbox{\hbox{$^1$}\kern-.1em $\slr$}}

\def\myprime{^\prime}
\pageheight{ 7.5 in}
\baselineskip=25pt plus 2pt
\nologo

%
%

\topmatter

\title
{Local geometric invariants}
{of integrable evolution equations}
\endtitle
\author
 Joel Langer  and  Ron Perline
\endauthor
\affil
Dept. of Mathematics, Case Western Reserve University \\
Dept. of Mathematics and Computer Science, Drexel University
\endaffil
\abstract
The integrable hierarchy of commuting vector fields
for the localized induction
\linebreak \linebreak
equation
of 3D hydrodynamics, and its associated recursion operator,
are used to generate \linebreak \linebreak
families of integrable evolution equations
which preserve local geometric invariants of the \linebreak
\linebreak
evolving curve
or swept-out surface.
$$ \quad $$
$$ \quad $$
\endabstract
\endtopmatter

\newpage

{\bf Introduction}.
The concept of the soliton and
the inverse scattering transform approach
to solving certain non-linear equations  were  introduced twenty-five years
ago in two landmark papers  by
Zabusky-Kruskal and Gardner-Greene-Kruskal-Miura [1,2]. Since then, soliton
theory, or the theory of integrable systems, has had enormous impact on
applied mathematics and mathematical physics. Water wave theory [3],
nonlinear optics [4], field theory [5], and relativity [6] are but a
few of the areas which have been influenced by these ideas. An important
aspect of the study of soliton equations is that a single example can be of
interest in a wide variety of contexts; this universality may well be
related to the fact that such equations frequently have an underlying
geometric meaning. For instance, the sine-Gordon equation, which first
appeared in differential geometry [7], is also used as a model for
dislocation of crystals [8], for field theory [9], and nonlinear optics
(self-induced transparency) [10]. In the same vein, several authors
[11], [12], [13],[14],[15],
[16], have developed the connection between the
Serret-Frenet equations and other elements of classical differential
geometry and various well-known integrable models, including the non-linear
Schr\"odinger equation, and the modified Korteweg-de Vries equation.

Continuing this geometric theme, we describe in this paper certain families of
integrable equations which govern the motion of curves in the plane, on the
sphere, and in three-dimensional space.  We focus our attention on evolution
equations which preserve distinguished
{\it local} invariants of the evolving space curve or associated swept-out
surface.

Our point of departure is the {\it localized induction equation} (LIE):
$\gamma_t = \gamma_s \times \gamma_{ss} = \kappa B$.   LIE
is an idealized
{\it local} model of the evolution of the centerline
of a thin vortex tube in a three dimensional inviscid incompressible fluid.
Here, the subscripts $t$ and $s$ denote partial differentiation with respect
to time $t$ and  arclength $s$ for the evolving curve $\gamma(s,t)$ in
$R^3$,  which
has curvature $\kappa$, torsion $\tau$, and Frenet frame $\{ T,N,B \}$;
the multiplication sign denotes cross product.
For derivation and history, we refer to [17],[18].
More accurate,  non-local models have  been considered;
we refer the reader  to the
work of Moore-Saffman [19], Klein-Majda [20] and the references mentioned
therein.
LIE has a second interpretation  as
a ``potential'' form for the
classical one-dimensional continuous Heisenberg ferromagnet [21].

Hasimoto [22]  uncovered the relation of  LIE  to soliton theory by showing
that
it induces an evolution on the {\it complex curvature}
$\psi = \kappa \, e^{ i \int^s {\tau \,  du}}$
governed by the {\it nonlinear Schr\"odinger equation} (NLS):
$\psi_t = i(\psi_{ss}  + {1 \over 2}|\psi|^2 \psi)$.
 NLS  is a well known example of an
equation with soliton solutions.
Since Hasimoto's discovery, the structure of  LIE  has been more
fully spelled out in the context of infinite dimensional Hamiltonian
systems.  In fact the Hamiltonian for  LIE  is just the arclength
functional on curves (as shown by Marsden-Weinstein [23], who also
described the relevant Poisson structure).  Further, associated to  LIE
is an infinite sequence of  commuting Hamiltonian vector fields,
with Hamiltonians which can be expressed as global geometric
invariants of the curves.  All of
these equations are of the form $\gamma_t = W = aT + bN + cB$, where
$W$ is a geometric vector field, which means the coefficients $a,b,c$ are
functions of $\kappa, \tau, \kappa\myprime = \kappa_s, \tau\myprime = \tau_s$ ,
and higher
derivatives with respect to $s$.  We list
the first few terms of the {\it localized induction hierarchy}
(LIH) [24]  of commuting vector fields, as well as their associated
Hamiltonians (the vector field $X_{0}$ is exceptional):
$$
\eqalign{
& X_{0} = -T,   \cr
& X_{1} = \kappa B, \quad  I_{1} = \int_\gamma \ ds , \cr
& X_{2} = {{\kappa^2} \over 2 }T  +   \kappa \myprime N + \kappa \tau B,
 \quad I_{2} = \int_\gamma -\tau \ ds ,\cr
& X_{3} \ =
\kappa^2 \tau T
+ (2 \kappa\myprime \tau + \kappa \tau \myprime)N
+ (\kappa \tau^2 - \kappa ^{\prime \prime} -{1 \over 2}\kappa^3)B ,  \cr
& \quad I_3 = \int_\gamma {1 \over 2} \kappa^2 \ ds , \cr
& X_{4} \ =
(-\k \k '' + {1 \over 2} (\k ')^2 + {3 \over 2} \k ^2 \t ^2 - {3 \over 8} \k
^4)T \cr
& + ( - \k ''' + 3 \k \t \t ' + 3 \k ' \t ^2 - {3 \over 2} \k ^2 \k ') N \cr
& + ( \k \t ^3 - 3 (\k ' \t ) ' -  {3 \over 2} \k ^3 \t - \k \t '') B, \cr
& \quad I_{4} = \int_\gamma  \, {1 \over 2} \kappa^2 \tau \,{ds}, \cr
& X_n \ = \slr(X_{n-1}), \quad  n > 0  \, .\cr
}
$$
As the last line suggests, LIH
is generated by
successively applying a recursion operator
starting with $X_{0}$.
For backround material on recursion operators in the theory of
integrable systems, see [25];  the recursion operator for LIH
was introduced in [24], and is defined  below.

The vector field $X_{2}$ also arises in a related fluid mechanical
context, via a refined version of  LIE.  Fukumoto-Miyazaki [26]
derived a ``localized induction equation"
for thin vortex tubes, which allows for axial velocity;
their equation be expressed in terms of the vector fields of LIH via
$  \gamma_t = ( cX_{2} + X_{1} )$,  $c$ some parameter.
Lamb [11] had previously considered the (essentially equivalent)  evolution
equation
$\gamma_t = X_{2} - 3 \tau_0 X_{1} + 3 {\tau_{0}}^2 X_{0}$,
which has the following special property:
starting with an initial curve $\gamma(s,0)$ having constant torsion
$\tau_0$, the evolution preserves the condition $\tau = \ \hbox{constant} \
= \tau_0$, while the curvature function $\kappa(s,t)$ satisfies the modified
Korteweg-deVries equation
$\kappa_t = \kappa_{sss} + {3 \over 2}\kappa^2 \kappa_s$, another well known
soliton equation.
Lamb's result can be interpreted as saying that there exist distinguished
solutions to the Fukumoto-Miyazaki equation which can be completely described
by tracking just one evolving functional parameter (curvature) rather
than the usual two (curvature and torsion).
The special case $\tau_0 =0$ corresponding to an evolution
of planar curves has been discussed in several recent papers [15],[16],[27];
in fact, in [28], this planar evolution equation is shown to be a "localized
induction approximation" associated with vortex patches in ideal
two-dimensional fluids.
Part of our aim here is to demonstrate that Lamb's example is  but one of a
whole family of evolution equations preserving local geometric invariants,
that such examples are geometrically significant, and that they are
intimately related to the recursion operator $\slr$.

{\bf The Recursion Operator $\slr$ and
Variation formulas}.
Formulas for the variations of geometric invariants of a curve
$\gamma(s,t)$ evolving by a vector field $\gamma_t = W$ have been
derived by various authors [16],[24],[29],[30].  Much of the present
paper depends on the remarkably simple expressions for these
variations in terms of the recursion operator $\slr$ in case
$W$ is {\it locally arclength preserving} (LAP), i.e., in case
$<W_s, T> = 0$.   In addition to curvature and torsion, we consider
{\it natural curvatures} $u$ and $v$ related to $\kappa$ and $\tau$
via $u + iv = \psi = \kappa e^{i \int^s \tau(u) du}$.  The corresponding
{\it natural frame} $\{ T,U,V \}$ is related to the standard
Frenet frame by $U + iV = (N + iB) e^{i \int^s \tau(u) du}$ and
satisfies the {\it natural Frenet equations}
$T_s = uU + vV, U_s = -uT, V_s = -vT$.  We begin our list of
formulas with the definitions of $\slr$ and a {\it parameterization operator}
$\slp$, which takes an arbitrary vector field
$X = aT + bN + cB = fT + gU + hV$ to an LAP vector field:

$$\leqalignno{
\slp(X) &= \int^s (\kappa b){\, ds \,} T + bN + cB &(a)  \cr
        &= \int^s (gu + hv) {\, ds  \, }T  + gU +hV,   \cr
\slr(W) &= -\slp( T \times W^\prime),  &(b) \cr
W(\kappa) &= \ <-\slr^2(W),N>, &(c) \cr
W(\tau) &= \ <-\slr^2(W),B/\kappa>' ,  &(d) \cr
W(u) &= \ <-\slr^2(W),U>, &(e) \cr
\quad  W(v) &= <-\slr^2(W),V> .  \cr
   }$$

Note that at each time $t$, $u+iv$ is only determined up to multiplication
by a complex unit; the formulas for $W(u)$ and $W(v)$ should therefore
read ``for some choice of natural curvatures $u(s,t)$ and $v(s,t)$".  Also,
the appropriate choice of antiderivative in the definition of $\slp$ depends
upon the class of curves under consideration.  In particular, in the
{\it asymptotically linear case}, it is the ``antisymmmetrized"
antiderivative operator $\int^s  f {\, ds \, } = F(s) -{1 \over 2}(F(\infty) + F(
-\infty)), F^\prime = f$, which gives the terms of LIH listed above.
The formulas stated above were derived in [24] and [31], where similar formulas
were also derived for the evolution of the frame $\{T,U,V\}$.

{\bf Curve invariants}.
We now present the integrable
hierarchies $X_n, Y_n$, $Z_n, \Omega_n$,
$\omega_n, \sigma_n, \Sigma_n$,
of vector fields preserving geometric invariants of
the evolving space curve $\gamma$.  As noted below, the $X_n$ and
$Y_n$ hierarchies have been discussed previously in the literature
(we include them for completeness); the other hierarchies
are new.

{\it (i) Locally arclength preserving:} For $n \geq 0$, the  LIE
hierarchy  $X_{n}$ is locally
arclength preserving. The $X_n$ satisfy the recursion relation
$X_{n+1} = \slr (X_n)$,  starting with $X_{0}$.
If $W = X_n$, then the induced evolution
on complex curvature $\psi_t = W(\psi)$ is the corresponding element
of the  NLS  hierarchy; in particular, $X_{1} = \kappa B$ induces the
the NLS equation (details appear in [24]).
Obviously, any linear combination of the $X_n$ will also be locally
arclength preserving.  Our hierarchies will mainly be of this form.

{\it (ii) planarity preserving:} For $n \ge 0$, the {\it even} flows
in  LIH  $ Y_n = X_{2n}$ are planarity preserving (the evolution of
a planar curve stays planar).  The $Y-$recursion operator is just
$\slr^2(W)= -\slp(W_{ss})$ restricted to planar vector fields
along planar curves.
The starting vector field in the sequence is $Y_0=X_{0}$.
If $W=Y_n$ then the induced evolution on curvature $\kappa_t = W(\kappa)$
is the corresponding element of the (mKdV) hierarchy; in particular,
$Y_{1}$ induces the (mKdV) equation itself [15],[16],[27].
Observe that planarity preserving is equivalent to preserving the
condition $\tau = 0$.

{\it (iii) constant torsion preserving:}  For $n \ge 0$, the
vector fields
$$Z_n = \sum_{k=0}^{2n+1} {{\binom {2n+1}k} (-\tau_0)^k X_{2n-k}}$$
preserve the constant torsion condition $\tau = \tau_0$.
Each
$Z_n$ contains a ``phantom" or formal vector field component $X_{-1}$; by
definition, $\slr(X_{-1})= X_{0}$.  $X_{-1}$ has trivial geometric
action;  it is introduced for algebraic purposes, to facilitate
the recursion formulas for the $Z_n$.  The $Z$-recursion operator
is $(\slr-\tau_0)^2$; the starting vector field
in the sequence is $Z_0 = X_{0}-\tau_0 X_{-1}$.
We remark that along constant torsion curves,
the operator $\slr - \tau_0$, thought of as an
integro-differential operator on vector fields $W = aT + bN + cB$,
has coefficients independent of $\tau$.
If $W=Z_n$ then the induced evolution
on curvature $\kappa_t=W(\kappa)$ is the corresponding element
of the (mKdV) hierarchy; in particular $Z_{1}$ induces the
(mKdV) equation, and we recover the result of Lamb [11].  When
$\tau_0= 0$, we are in case (ii).

A closely related sequence of vector fields is given by
the ``planar-like" vector fields
$$\Omega_n = \sum_{k=0}^{2n-1} {{\binom {2n-1}k} (-\tau_0)^k X_{2n-k}} ,$$
so called because they have no binormal component
and their coefficients are independent of $\tau$;
in fact, the $\Omega_n$ look identical to the $Y_n$,
the only difference being that they are defined  along
space curves rather than planar curves.    We have
$\Omega_{n+1} = \slr^2 Z_n$.  Again, the recursion
operator is $(\slr - \tau_0)^2$ and the starting vector field
in the sequence is $\Omega_1=X_{2} - \tau_0 X_{1} =
{{\kappa^2} \over 2}T + \kappa^{\prime}N$.  The form of $\Omega_n$,
together with formulas (c) and (d),
imply that the $Z_n$ are constant torsion preserving and induce the
(mKdV) hierarchy on the curvature $\kappa$.

{\it (iv) Constant natural curvature and sphericity preserving:}
Note that if a natural curvature is constant ($v = v_0$ along a curve
$\gamma$) then the  natural Frenet equations imply $c = \gamma + (1/v)V$ is
a constant vector, so $\gamma$ lies on a sphere of radius $1/v$ centered
at $c$.  (Compare this with the standard sphericity condition
$\kappa^{-2} + (\kappa^\prime)^2 \tau^{-2} \kappa^{-4} = \hbox{constant},$
given in most differential geometry books, e.g., [32]).  Conversely,
if $\gamma$ lies on a sphere of of radius $r$, then a natural frame
$\{T,U,V\}$ along $\gamma$ is readily obtained by letting $U$ be normal
to $T$ and tangent to the sphere (and so $V = T \times U$  is normal to the
sphere); then the corresponding natural curvature $v$ is constant,$v=1/r$,
and $u$ is the {\it geodesic curvature} of $\gamma$, regarded as a
spherical curve.

We define a constant $v$-preserving analogue of the planar hierarchy
$Y_n$: $\omega_1 = {{u^2} \over 2}T + u_s U$, etc., are obtained by
simply replacing $\kappa$ by $u$ and $N$ by $U$ in the expressions for
the corresponding $Y_n$.  From this description, it is clear that the
$\omega_n$ define evolutions on spherical curves; further, they are
generated by an {\it intrinsic recursion operator}, $\sls(W) =
- \slp (\nabla_T^2(W))$ --  here $\nabla$ is
covariant differentiation on the sphere --
since $\nabla_T(T) = uU$, and $\nabla_T(U) = -uT$ look just like the
planar Frenet equations.  In terms of $\slr$, we have $\sls(W) =
(\slr^2 + v^2)(W)$; specifically, if we consider the class of
{\it asymptotically geodesic curves} -- $u$ decays rapidly to zero --
the antiderivative operators in both formulas for  $\sls$ are the
``antisymmetrized" ones, mentioned above.

We note that, just as Euclidean elastic curves give solitons for
LIE evolving by screw motion, ``spherical elastic curves" [33]
simply  rotate when evolving via $\gamma_t = \omega_1$.  For a
general spherical curve, one might expect $u$ to evolve according
to mKdV, by analogy with the planar case. This is not quite true; however,
the related spherical vector fields defined by
$\sigma_1 = ({{u^2} \over 2} -
v^2)T + u_sU, \,  \sigma_{n+1} = \sls \sigma_{n}$
induce exactly the mKdV hierarchy (and this links our work with
that of Chern-Tenenblat [34] on integrable foliations of surfaces
with constant Gauss curvature). The two families are related via
$\omega_{n+1} = \slr^2(\sigma_n)$.

Yet another hierarchy of spherical evolution equations is obtained
by simply restricting the even  vector fields $X_{2n}$ to curves with
$v = v_0$.  It is worth remarking that both the $\sigma_n$ and $\omega_n$
can be expressed in terms of the restricted $X_{2n}$:
$$\eqalign{
\omega_n &= \sum_{k=0}^{n}
{ {(-1)^k v_0^{2k} ({1 \over 2}-n)_k} \over {k!} }  X_{2n-2k} \, ,  \cr
\sigma_n &= \sum_{k=0}^{n}
{{(-1)^k v_0^{2k} (-{{1}\over 2}-n)_k} \over {k!}}  X_{2n-2k} \cr
}
$$
Here, $(a)_k = (a)(a+1)(a+2) \dots (a+k-1)$ is the standard
Pochhammer symbol.

{\it (v) Constant curvature preserving}: The interest of the spherical
evolution equations discussed above
is enhanced by the general connection between
spherical curves and space curves, via the {\it tangent indicatrix}
construction.  To be brief, we mention only the following
special case: a unit speed curve $\gamma$ on the unit sphere, with
geodesic curvature $u$, is the derivative of a unit-speed space curve $\Gamma$,
having constant curvature $\kappa=1$ and torsion $\tau=u$.  It follows
that we may regard the spherical equations of (iv) as evolution equations
on space curves of constant curvature $\kappa=1$; moreover, in the case
of the $\sigma_n$ hierarchy, the torsion of $\Gamma$ evolves according
to mKdV.  We note that it is also possible to translate the $\sigma_n$ into
explicit (non-local) vector fields on the space curve level; e.g.,
$\sigma_1$ corresponds to $\Sigma_1 = \int^s
{(
({{\tau^2} \over 2} -1)N
+\tau_s B \,) \,  ds}$, etc.

{\bf Surface invariants}.   Given an evolving space curve, one can also
consider the geometric invariants of the {\it surface} being swept out
by the curve.  For example, in case (iv), the curves are sweeping out a
sphere, a surface of  constant positive Gauss curvature.

We now discuss means of generating  constant negative Gaussian curvature
(CNGC) surfaces.  These surfaces
have long been a subject of interest in differential geometry, from
the early work of  B\"acklund [35],
to the recent work of Sym [36] and
Melko-Sterling [37], who have approached their
study using  some of the tools of
modern soliton theory.
Here, we indicate a
complementary avenue of investigation: we discuss evolution equations
on curves which sweep out CNGC surfaces.

We begin by observing that the variation formulas
$(c), (d), (e)$ clearly simplify in the case
that $W$ is an {\it eigenvector} for $\slr$.
Specifically, if $\gamma$ is a curve with constant
torsion $\tau = \tau_0$, one easily verifies that
the vector field
$W = cos(\theta) T - sin(\theta) N$
satisfies $\slr (W) = \tau_0 W$,
where $\theta = \int^s \kappa(u) du$.
Note that as a consequence of our variation formulas,
the evolution equation $\gamma_t = W$ induces an evolution on
curvature
$\kappa_t = \, -<N,\slr^2(W)> \, = \,
 -<N,\tau_0^2 W >
\, = \tau_0^2 sin(\theta)$,
so  $\theta$ satisfies the {\it sine-Gordon equation}
$$\theta_{st} = -G sin(\theta), \quad \hbox{where} \ G = -\tau_0^2  .$$
This well-known soliton equation has long been associated with
CNGC surfaces.  Let us substantiate the claim that the swept out surface
$M$ is indeed CNGC, with curvature $G$.
Observe, via our variation formulas, that $W(\tau) = 0$.  Thus constant
torsion is preserved during  the evolution.
Next, the normal $\nu$ to the surface is just
the binormal $B$ of our evolving curve; hence the {\it second fundamental form}
of the surface,
applied to the tangent $T$ of our curve $\gamma$,
is just $<-d\nu(T),T>  \myequal  <-B_s, T>
\myequal  <\tau N, T> \myequal 0$.
Thus $T$ is an {\it asymptotic direction} of the surface $M$.
By the Beltrami-Enneper theorem [38],
the Gauss curvature of the surface $M$ along
$\gamma$ is  the negative of the square of the torsion of $\gamma$;
this is just
$G$.
This {\it dynamical} prescription for generating CNGC surfaces,
although closely related
to standard discussions of the sine-Gordon equation [11],[16],
seems to be new (it was recently and independently derived
by Segur and McLachlan [39] in their  study of evolution equations
associated with surface motion in $R^3$).

Let's consider this evolution equation
$\gamma_t = W$
when the initial curve is
a {\it Hasimoto filament} [22], a curve $\gamma$ with curvature
$\kappa= 2a  \,\hbox{sech}(as)$ and torsion $\tau = \tau_0$.  One easily
checks (with appropriate choice of antiderivative $\theta = \int^s \kappa ds$ )
that $\theta_{ss} = a^2 sin(\theta)$. Thus $W = cos(\theta)T - sin(\theta)N
= -a^{-2}( ({{\kappa^2} \over 2} -a^2 )T + \kappa^{\prime}N)$.  Consequently,
modulo
scaling and slippage along the curve, $W = \Omega_1$, one of our previous
planar-like vector fields.

Along $\gamma$,  we have $\Omega_1 = X_2 - \tau_0 X_1$.  Both $X_2$ and
$X_1$ are {\it Killing fields} (= infinitesimal isometries) along
$\gamma$ [29]; hence the curvature and torsion functions remain unchanged
(modulo translation) as the curve evolves.  The comments of the previous
paragraph therefore apply for all time; and the result is that the
Hasimoto filament, evolving via the
equation $\gamma_t = \Omega_1$,
sweeps out a CNGC surface.
We contrast this result with
an observation of Melko-Sterling:  under the evolution
$\gamma_t = X_1$, the Hasimoto filament sweeps out a
surface $M$ which is a  {\it formal boundary point }  of
the set of CNGC surfaces, although $M$ itself is not CNGC.
As our simple example suggests, there is a connection between
the hierarchies of vector fields described in this paper, and
the geometry
of CNGC surfaces  --
the vector fields $\Omega_n$ play a central
role.
Details will appear in a subsequent
publication.

\end
\newpage

\Refs\nofrills{References}


\widestnumber\key{Ge-Di23}


\ref \key {\bf 1}  \by N.J. Zabusky and M.D. Kruskal
\jour Phys. Rev. Lett.
\vol 15
\yr 1965
\page 240
\endref  \ms \ms

\ref \key {\bf 2} \by C.S. Gardner, J.M. Greene, M.D. Kruskal and R.M. Miura
\jour Phys. Rev. Lett.
\vol 19
\yr 1967
\page 1095
\endref  \ms \ms


\ref \key {\bf 3} \by H. Hasimoto and H. Ono
\jour J. Phys. Soc. Japan
\yr 1972
\page 805
\endref  \ms \ms

\ref \key {\bf 4} \by A. Newell and J. Moloney
\book Nonlinear optics
\publ Addison-Wesley
\publaddr Redwood City, CA
\yr 1992
\endref  \ms \ms

\ref \key {\bf 5} \by L. L. Chau
\book Proceedings of the Thirteenth International
Colloquium of Group Theoretical
\linebreak
\quad \quad \quad
\linebreak
Methods in Physics
\publ World Scientific
\publaddr Singapore
\yr 1984
\endref  \ms \ms


\ref \key {\bf 6} \by V.A. Belinskii and V.E. Zakharov
\jour Sov. Phys. JETP
\yr 1978
\vol 48
\page 985
\endref  \ms \ms

\ref \key {\bf 7} \by L.P. Eisenhart
\book A treatise of the differential geometry of curves
and surfaces
\publ Dover
\yr 1960
\endref  \ms \ms


\ref \key {\bf 8} \by A. Seeger, H. Donth and A. Kochenforder
\jour Z. Phys.
\yr 1953
\vol 134
\page 173
\endref  \ms \ms

\ref \key {\bf 9} \by J.K. Perring and H.R. Skyrme
\jour Nucl. Phys.
\vol 32
\yr 1962
\page 550
\endref  \ms \ms

\ref \key {\bf 10} \by S.L. McCall and E.L. Hahn
\jour  Phys. Rev. Lett.
\vol 18
\page 908
\yr 1967
\endref  \ms \ms


\ref \key{\bf 11} \by G.L. Lamb, Jr.
\jour Phys. Rev. Lett.
\yr 1976
\vol 37(5)
\page 235
\endref  \ms \ms

\ref \key{\bf 12 } \by M. Lakshmanan
\jour J. Math. Phys.
\vol 20(8)
\yr 1979
\page 1667
\endref  \ms \ms

\ref \key{\bf 13} \by G. Reiter
\jour J. Math. Phys.
\vol 21(12)
\yr 1980
\page 2704
\endref  \ms \ms


\ref \key{\bf 14} \by  J. Cieslinski, P. Gragert, A. Sym
\jour Phys. Rev. Lett.
\yr 1986 \vol 57 \page 1507
\endref  \ms \ms


\ref \key{\bf 15} \by R. Goldstein and D. Petrich
\jour Phys. Rev. Lett.
\yr 1991 \vol 67 \page 3203
\endref  \ms \ms

\ref \key{\bf 16} \by K. Nakayama, H. Segur, and M. Wadati
\jour Phys. Rev. Lett.
\vol 69 \yr 1992 \page 2603
\endref  \ms \ms


\ref \key{\bf 17} \by  G.K. Batchelor
\book  An introduction to fluid dynamics \publ Cambridge
University Press \publaddr
\linebreak
\quad \quad \quad
\linebreak
 New York  \yr1967
\endref  \ms \ms

\ref \key{\bf 18} \by R. Ricca
\jour Nature \yr 1991 \vol 352 \page 561
\endref  \ms \ms

\ref \key{\bf 19} \by  D. Moore and P. Saffman
\jour Phil. Trans. R. Soc. London
\vol A 272 \yr 1972 \page 403
\endref  \ms \ms

\ref \key{\bf 20} \by  R. Klein and A. Majda
\jour Physica D
\yr 1991 \vol 49 \page 323
\endref  \ms \ms


\ref \key {\bf 21} \by M. Lakshmanan
\jour  Phys. Lett.
\vol 61A
\yr 1977
\page 53
\endref  \ms \ms




\ref \key{\bf 22} \by  H. Hasimoto
\jour  J. Fluid Mech.
\yr1972 \vol 51 \page 477
\endref  \ms \ms

\ref \key{\bf 23} \by  J. Marsden and A. Weinstein
\jour Physica 7D \yr1983 \page 305
\endref  \ms \ms

\ref \key{\bf 24}
\by  J.Langer and R. Perline
\jour J. Nonlinear Sci.
\yr 1991 \vol 1 \page 71
\endref  \ms \ms

\ref \key{\bf 25} \by P.J. Olver
\jour Proceedings of the International School on Applied Mathematics,Paipa,
Columbia
\endref  \ms \ms

\ref \key{\bf 26} \by  Y. Fukumoto and T. Miyazaki
\jour J. Phys. Soc. Japan
\yr 1986 \vol 55 \page 4152
\endref  \ms \ms



\ref \key{\bf 27}
\by J. Langer and R. Perline
\paperinfo submitted to the Fields Institute Proceedings:
Mechanics Days
\linebreak
\quad \quad \quad
\linebreak
Workshop, June 1992
\endref  \ms \ms


\ref \key{\bf 28} \by R. Goldstein and D. Petrich
\jour Phys. Rev. Lett.
\yr 1992 \vol 69 \page 555
\endref \ms \ms


\ref \key{\bf 29} \by J.Langer and D. Singer
\jour  J. London Math. Soc.(2)
\yr1984
\vol 30 \page 512
\endref  \ms \ms


\ref \key{\bf 30} \by S. Langer and R. Goldstein
\paperinfo (to be published)
\endref  \ms \ms

\ref \key{\bf 31}
\by J. Langer and R. Perline
\paperinfo submitted to the Fields Institute Proceedings:
Mechanics Days
\linebreak
\quad \quad \quad
\linebreak
Workshop, June 1992
\endref  \ms \ms


\ref \key{\bf 32} \by D. Struik
\book Lectures on classical differential geometry
\publ Dover Publ.
\publaddr New York \yr 1988
\endref  \ms \ms

\ref \key{\bf 33} \by  J. Langer and D. Singer
\jour J. Diff. Geom
\yr1984 \vol 20 \page 1
\endref  \ms \ms

\ref \key{\bf 34} \by  S.S. Chern and K. Tenenblat
\jour J. Diff. Geom
\vol 16
\yr 1981
\page 347
\endref  \ms \ms


\ref \key{\bf 35} \by A. B\"acklund
\jour Math.  Ann.
\vol 19
\yr 1882
\page  387
\endref  \ms \ms


\ref \key{\bf 36} \by  A. Sym
\paperinfo {\it
Geometrical aspects of the Einstein equations and
integrable systems}, vol. 239, \linebreak
\newline
Lecture Notes in Physics, 1985, p. 154
\endref \ms \ms



\ref \key{\bf 37} \by M. Melko and I. Sterling
%
\jour Ann. of Glob. Anal
\paperinfo  to appear
\endref  \ms \ms



\ref \key {\bf 38} \by M. Spivak
\book A comprehensive introduction to differential geometry
\publ Publish or Perish
\publaddr Houston $$ \quad $$
\yr 1979
\endref  \ms \ms

\ref \key{\bf 39} \by R. McLachlan and H. Segur
%
\paperinfo preprint
\endref \ms \ms


\endRefs
